%% file: main.tex
\documentclass[acmsmall]{acmart}
\AtBeginDocument{%
  \providecommand\BibTeX{{%
    \normalfont B\kern-0.5em{\scshape i\kern-0.25em b}\kern-0.8em\TeX}}}

\setcopyright{acmcopyright}
\copyrightyear{2018}
\acmYear{2018}
\acmDOI{XXXXXXX.XXXXXXX}

\acmConference[Conference acronym 'XX]{Make sure to enter the correct
  conference title from your rights confirmation emai}{June 03--05,
  2018}{Woodstock, NY}
\acmBooktitle{Woodstock '18: ACM Symposium on Neural Gaze Detection,
 June 03--05, 2018, Woodstock, NY} 
\acmPrice{15.00}
\acmISBN{978-1-4503-XXXX-X/18/06}

\usepackage{subfiles}
\usepackage{url}
\usepackage{subcaption}
\usepackage{xspace}
\usepackage{xpunctuate}
\usepackage{enumitem}
\usepackage{wrapfig}
\usepackage{graphicx}
\usepackage{indentfirst}
\usepackage{soul}
\usepackage{ulem}

\newcommand{\ziwei}[1]{{\color{teal}{ (Ziwei: #1)}}} 

\newcommand{\remove}[1]{\textcolor{red}{\sout{#1}}}
\renewcommand{\remove}[1]{\unskip}
\newcommand{\revise}[1]{\textcolor{black}{#1}}

\definecolor{bct}{RGB}{0,0,0}
\definecolor{moa}{RGB}{0,0,0}

\definecolor{deepskyblue}{rgb}{0.0, 0.0, 0.0}

\newcommand{\fbc}[1]{{\color{deepskyblue}{\textit{#1}}}}    
\newcommand{\bct}[1]{{\color{bct}{\textit{#1}}}}    
\newcommand{\moa}[1]{{\color{black}{\textit{#1}}}}    

\newcommand{\ie}{{i.e.,}\xspace}
\newcommand{\eg}{{e.g.,}\xspace}
\newcommand{\etal}{{et~al\xperiod}\xspace}

\newcommand{\etc}{{etc\xperiod}\xspace}

\begin{document}

\title{Behavior Matters: An Alternative Perspective on Promoting Responsible Data Science}

\author{Ziwei Dong}
\email{ziwei.dong@emory.edu}
\affiliation{
    \institution{Emory University}
    \city{Atlanta}
    \state{Georgia}
    \country{USA}
}

\author{Ameya Patil}
\email{ameyap2@cs.washington.edu}
\affiliation{
    \institution{University of Washington}
    \city{Seattle}
    \state{Washington}
    \country{USA}
}

\author{Yuichi Shoda}
\email{yshoda@uw.edu}
\affiliation{
    \institution{University of Washington}
    \city{Seattle}
    \state{Washington}
    \country{USA}
}

\author{Leilani Battle}
\email{leibatt@cs.washington.edu}
\affiliation{
    \institution{University of Washington}
    \city{Seattle}
    \state{Washington}
    \country{USA}
}

\author{Emily Wall}
\email{emily.wall@emory.edu}
\affiliation{
    \institution{Emory University}
    \city{Atlanta}
    \state{Georgia}
    \country{USA}
}
\begin{abstract}
  Data science pipelines inform and influence many daily decisions, from what we buy to who we work for and even where we live. When designed incorrectly, these pipelines can easily propagate social inequity and harm. Traditional solutions are technical in nature; e.g., mitigating biased algorithms. In this vision paper, we introduce a novel lens for promoting responsible data science using theories of behavior change that emphasize not only technical solutions but also the behavioral responsibility of practitioners. By integrating behavior change theories from cognitive psychology with data science workflow knowledge and ethics guidelines, we present a new perspective on responsible data science. We present example data science interventions in machine learning and visual data analysis, contextualized in behavior change theories that could be implemented to interrupt and redirect potentially suboptimal or negligent practices while reinforcing ethically conscious behaviors. We conclude with a call to action to our community to explore this new research area of behavior change interventions for responsible data science.
\end{abstract}



     %

\maketitle



\subfile{sections/1.introduction}

\subfile{sections/2.behavior_change}

\subfile{sections/3.ds}

\subfile{sections/operationalize}

\subfile{sections/4.interventions_14_sept}
\subfile{sections/5.discussion}

\subfile{sections/6.conclusion}

\bibliographystyle{ACM-Reference-Format}
\bibliography{sample-base,behavior_change}

\appendix

\end{document}

%% file: sections/1.introduction.tex
\section{Introduction}

While data science can advance important societal goals, such as fighting climate change and species extinction, it can also cause considerable societal harm~\cite{barocas2017engaging}. 
Individual mispredictions can lead to the dehumanization of Black people by labeling them as gorillas~\cite{mulshine2015major}, or loss of health benefits for those who need them the most~\cite{eubanks2018automating}.
These examples hint towards larger systems of inequity that data science pipelines inadvertently perpetuate when left unchecked.

We have seen a heartening surge in academic research to counteract these inequities in machine learning and broader data science practices~\cite{mehrabi2022survey,lin2023bias,piorkowski2023aimee,eardley2023explanation}, including the introduction of the Conference on Fairness, Accountability, and Transparency in 2018, and numerous workshops such as 
Community-Driven AI: Empowering People Through Responsible Data-Driven Decision-Making at CSCW 2023. Existing work often focuses on
ensuring that the data science pipelines, and consequently their outputs, are mathematically and statistically sound. Issues of bias and inequity are then framed as mitigating erosion of technical quality, such as detecting and counteracting biased input data or biased algorithms; for example, developing bias mitigation strategies to counter bias in face detection datasets~\cite{yang2022enhancing,buolamwini2018gender}.

However, modifying the algorithms and models that data scientists use is not enough to solve such a systemic problem. 
We liken this solution to modifying cigarettes to prevent lung cancer rather than helping smokers quit smoking. A technical solution may be satisfactory for avoiding traditional cigarettes, but it does not help people avoid addictive behaviors.
Similarly, while we observe that technical solutions are essential to successful data science we also argue that they are insufficient for ensuring responsible outcomes in human-AI interactions. Biases appear within datasets and algorithms because \emph{people} inadvertently put them there. When we focus on the \textit{inputs} (data, algorithms) and \textit{outputs} (inferences) and not on the \textit{agents} involved (people and systems), we may miss the opportunity to more meaningfully address the underlying causes of the problems we seek to fix. 

The CSCW \remove{and CHI} community has addressed the responsibility of the \textit{agents} involved through exploring and understanding human factors in responsible machine learning and model interpretability\cite{hong2020human}, the influence of human interaction on the efficacy of ML model-driven decision-making\cite{green2019principles}, and dissonance in the perception of human and machine understanding\cite{zhang2019dissonance}. \remove{but this alternative viewpoint has not been formally studied yet.} 
\revise{Collaborative data science for social good is of particular interest to the CSCW community~\cite{zhang2019dissonance,green2019principles,hong2020human,zhang2020data,passi2018trust,wang2019human,meng2019collaborative,tomavsev2020ai}.
Our work complements these existing efforts by focusing on the human behaviors and interactions that influence responsible data science in both individual and collaborative data science settings. First, data science projects have a broad impact on society and communities\cite{meng2019collaborative,tomavsev2020ai}. Encouraging responsible data science practices can promote social good at large. Second, data science itself is a community where practitioners must collaborate to deliver responsible models\cite{zhang2020data,passi2018trust}. Ensuring that data science practitioners follow the same responsible behavior principles fosters collaboration and facilitates the effective delivery of data science projects\cite{wang2019human}. By integrating behavior change theories with data science practices, we offer novel methods to support ethical decision-making and collaborative efforts in data science, ultimately contributing to the design, development, and analysis of computer-supported collaborative systems.}

\begin{figure}
  \includegraphics[width=\textwidth, trim={0 0cm 0 0cm}, clip]{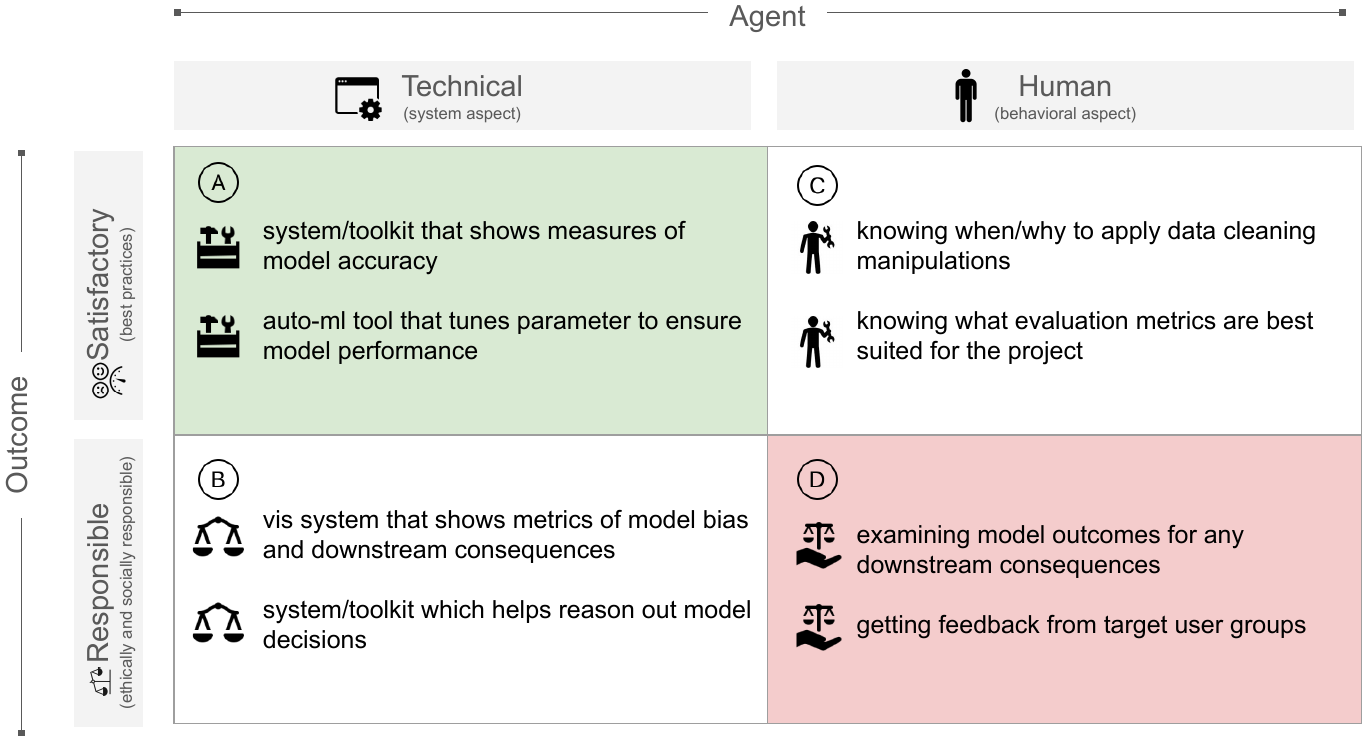}
  \caption{
    We characterize data science practices according to desired outcomes (rows -- satisfactory and responsible) and agents (columns -- technical and human). \revise{It is important to note that outcomes are not mutually exclusive.} Rigorous data science has historically emphasized technical aspects like auto-tuning and measures of model accuracy (A, green cell). Recent efforts towards model fairness have illustrated responsible data science, but still ultimately rely on technical indicators and algorithmic solutions (B). In this paper, we emphasize the agency of humans (C and D, right-hand column), and in particular, how human behaviors can contribute to responsible data science (D, red cell). 
  }
  \Description{A 2x2 matrix for characterizing behaviors in data science}
  \label{fig:teaser}
\end{figure}

\textbf{In this vision paper, we explore opportunities to formally redefine responsible data science to encompass not only technical responsibility (holding algorithms/datasets accountable) but also behavioral responsibility, i.e., holding data scientists accountable for the \emph{patterns of behavior} that may lead to positive or negative social outcomes}. To this end, we reframe existing literature on data science best practices and ethics guidelines through the lens of behavior change models\remove{ from cognitive and clinical psychology}. To ground our discussion, we draw parallels from successful behavior change interventions from cognitive and clinical psychology such as smoking cessation~\cite{borrelli1994goal} to common data science scenarios today such as model training and exploratory visual data analysis (see \autoref{fig:examples}). In this way, we show how several key principles from foundational behavior change research translate to the data science domain.

Although there have been extensive studies on behavior change approaches~\cite{pinder2018digital} and data science ethics~\cite{aragon2022human} separately, we recognize that relatively few projects make \textit{direct} contributions at this intersection of behavior change interventions and data science. Our work culminates in a framework that can help researchers and designers navigate the vast space of prior work relevant to behavior change interventions and apply it in data science contexts. Thus we assert that our work serves as a foundational \textbf{call-to-action}: to inspire a \emph{critical research agenda} focused on cultivating responsible users beyond traditional technical education and training contexts; specifically,
\emph{through everyday interactions with data science tools}. \revise{Our objective is to foster reflection among data scientists regarding the significance of their actions towards practicing responsible data science.} Finally, we connect our vision to the larger effort to make data science more equitable and just, and outline open challenges for the community moving forward. To summarize, we make the following contributions:

\begin{itemize}
    \item We introduce the concept of \textit{behavior change interventions for data science}, where we focus on data science behaviors as possible predictors of biased outcomes. 
    
    \item In \autoref{sec:bc_psych}, we illustrate how existing psychological models can be applied to inform the design of behavior change interventions in data science. We further introduce how to operationalize them in \autoref{sec:operationalize}. 
    
    \item In \autoref{sec:background_ds}, we synthesize a definition of \emph{responsible} practices in data science work for humans and systems.
    
    \item In \autoref{sec:interventions}, we present concrete examples of possible interventions to encourage responsible practices within the data science context.
    
    \item We conclude with a discussion of open challenges in the space of behavior change interventions for responsible data science in~\autoref{sec:discussion}.
    
\end{itemize}

For reference, we include Table~\ref{table:acronyms} below to summarize acronyms that will be introduced and used throughout the remainder of the paper.

\subsection{Example Behavior Change Contexts}
\label{sec:examples}

\begin{figure}
    \centering
    \includegraphics[width=\textwidth]{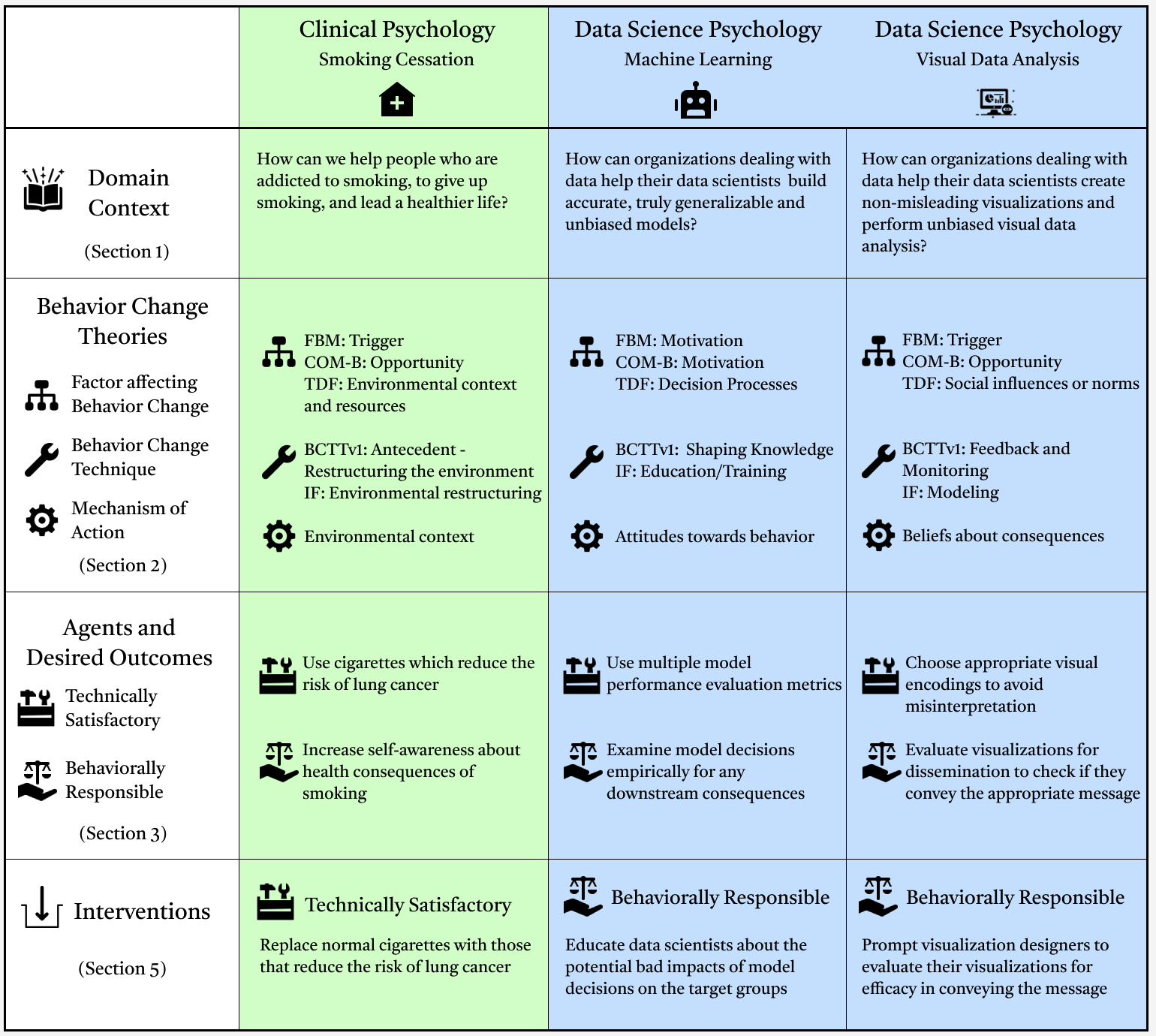}
    \caption{Drawing analogies from behavior change solutions in the clinical domain (green) to the data science domain (blue). Each column represents a behavior change domain. The rows characterize the behavior change problem \remove{, requirements,} and solutions\revise{, starting with the domain context.
    The next row characterizes exemplary theories of behavior change, followed by Agents and Desired Outcomes, and how together these might inform a specific intervention in each domain context (final row). The agents and outcomes, characterized as technically satisfactory or behaviorally responsible, are described further in Figure~\ref{fig:teaser} and Section~\ref{sec:outcomes_agents}.}
    \remove{Note that in row 3 we only show two out of the four possible agent + desired outcome combinations as per~\autoref{fig:teaser} -- \textit{technically satisfactory} which is what data science has historically focused on, and \textit{behaviorally responsible} which is the focus of our work. Further, we show interventions (row 4) and the corresponding behavior change theories (row 2) for one of the two agents and desired outcome combinations illustrated in row 3.} We hand-pick these limited examples for the sake of space and to demonstrate how behavior change theory can be applied across different domains to bring about the desired outcome through the agent in a generalizable way. 
    }
    \label{fig:examples}
\end{figure}

To ground our forthcoming discussion on behavior change theories, we introduce three concrete examples, 
shown in \autoref{fig:examples}. 
Each example is described according to the context of the domain and the agent and desired outcome (as shown in Figure~\ref{fig:teaser}). 
The figure also details the behavior change theories relevant to these contexts and example behavior change interventions, which we will introduce in Sections~\ref{sec:bc_psych} and \ref{sec:interventions}, respectively. 

The first domain (green column) represents a context in which behavior change interventions have been previously employed. 
In clinical psychology, smoking cessation is a well-known problem that has been approached from numerous perspectives, including that of behavior change~\cite{borrelli1994goal, paay2015understanding}. 
In this case, researchers and clinicians are interested in how to help people live a healthier life by giving up smoking. 

The last two domains (blue columns) represent our vision for how we can design behavior change interventions in data science contexts using existing behavior change theories. 
We cover two example tasks. 
First, we consider a machine learning context, wherein organizations \remove{are interested in helping} \revise{encourage} their data scientists build accurate, generalizable, and unbiased models. Finally, we consider a visual data analysis context, wherein stakeholders want to ensure that visual data analysis practices and resulting communications are fair and accurate. We use these three example contexts as a running example throughout the paper to review existing behavior change theories, and also to apply the framework provided by these theories for designing interventions for responsible data science.

\begin{table}[]
\caption{A summary of the acronyms used throughout this paper.\vspace{-2mm}}
\label{table:acronyms}
\begin{tabular}{@{}ll@{}}
\toprule
\textbf{Acronym} & \textbf{Meaning}                  \\ \midrule
\textbf{FBC}     & Factors affecting Behavior Change \\
\textbf{BCT}     & Behavior Change Techniques        \\
\textbf{MoA}     & Mechanisms of Action              \\
\textbf{FBM}     & Fogg Behavior Model \cite{fogg}\\
\textbf{COM-B}     & Capability (C), Opportunity (O), and Motivation (M) - Behavior (B) \cite{michie2011behaviour}\\
\textbf{TDF}     & Theoretical Domains Framework \cite{michie2005making, atkins2017guide} \\
\textbf{BCTT}     & Behavior
Change Techniques Taxonomy \cite{abraham2008taxonomy, michie2013behavior}             \\ \bottomrule
\end{tabular}
\end{table}

%% file: sections/2.behavior_change.tex
\section{Identifying Relevant Theories of Behavior Change for \remove{Responsible} Data Science}
\label{sec:bc_psych}

In order to deliver responsible behaviors in data science, we seek to understand the heuristics behind effective behavior change techniques and transfer them into the data science domain. In this section, we illustrate how existing psychological models can be applied to inform the design of behavior change interventions in data science. 

There have been numerous applications of behavior change techniques in the space of personal health such as for smoking cessation~\cite{borrelli1994goal, paay2015understanding}, in environmental domains such as for managing carbon footprint~\cite{nielsen2020behavioral, rau2022systematic}. 
In spite of the application of behavior change interventions in numerous domains, a survey by Wiafe~\etal~\cite{wiafe2012bibliographic} revealed that only half of the behavior change interventions in persuasive systems across the domains of health, commerce, education, and environment have a theoretical grounding. Orji~\etal~\cite{orji2018persuasive} revealed a similar finding for persuasive technologies in the clinical domain. Furthermore, prior works suggest that behavior change interventions informed by psychology theory are more effective than those that are not~\cite{michie2005making, cane2012validation}, promoting what is known as \emph{evidence-based practices}. Accordingly, there has been substantial theoretical development on evidence-based behavior change interventions. 

\revise{To identify relevant behavior change theories, we conducted a literature search beginning with two canonical theories on factors influencing behavior change by Fogg \cite{fogg} and Michie et al. \cite{michie2011behaviour}. 
We collected relevant papers by searching forward- and back-references, as well as conducting additional keyword-based searches in Google Scholar, including keywords such as ``behavior change theories'' and ``behavior change interventions.''
Throughout this exploratory process, we prioritized selecting theories and studies that are not only highly cited but have also stood the test of time (i.e., are still cited by a significant body of research at the time of this writing). 
From this corpus of relevant theories, we then grouped them into the following three categories: 
}
\remove{These theories of behavior change can be broadly classified into 3 categories:}

\begin{enumerate}
\item \textbf{Factors Affecting Behavior Change (FBC)} which tell us about the individual or group-level characteristics that can influence the likelihood of a target behavior being achieved, 
\item \textbf{Behavior Change Techniques (BCT)} which are specific techniques or interventions that, leveraging particular factors, can increase the likelihood of a target behavior, and 
\item \textbf{Mechanisms of Action (MoA)} which explain the underlying cognitive mechanism that makes a specific factor or technique work to influence behavior.
\end{enumerate}

That being said, many theories in each category have overlapping constructs~\cite{fishbein2001factors, michie2005making, atkins2017guide, cane2012validation, michie2013behavior, abraham2008taxonomy} which have been shown to make it difficult to identify individual processes or factors underlying successful behavior change~\cite{orji2018persuasive}. Further, most of these theories are rooted in psychology and clinical research with limited empirically verified attempts at generalization across different fields. Thus, rather than comprehensively surveying theories of behavior change in this paper, we instead focus on identifying and discussing the theories that appear most relevant for the data science context, as advised by Pinder~\etal~\cite{pinder2018digital} and Michie~\etal~\cite{michie2005making}. We accordingly choose theories that are highly cited and have more tangible implementations. 
\remove{We exclude theories of behavior change from our review that are too specific to the application domain \cite{fishbein2005theory, steinmetz2016effective, wantland2004effectiveness} and defer theories designed primarily for longer-term intervention \cite{fjeldsoe2009behavior, steinmetz2016effective} to the discussion (Section~\ref{sec:discussion-habit}). Taking together theories that grapple with factors affecting behavior change, behavior change techniques, and mechanisms of action, we form a strong theoretical foundation from which to envision possible interventions to promote ethical data science practices.} In this section, we describe these theories and use them to characterize behavior change in the domain contexts listed in~\autoref{fig:examples} to ground them.
We \remove{later} demonstrate how \remove{we can} \revise{to} use these theories to generate a series of interventions in \autoref{sec:interventions}.

\subsection{Factors Affecting Behavior Change (FBC)} 
\label{sec:3.1}
Here, we summarize established theories describing key factors
that influence behavior change. While certainly not exhaustive, we focus on the following three prominent theories because they are well-established in the literature and complementary within the context of data science.

    \begin{enumerate}

        \item \emph{\textbf{Fogg Behavior Model}}~\cite{fogg} or FBM (Fogg Behavior Model) proposes that behavior is comprised of three primary components: \textit{motivation, ability}, and \textit{trigger}. \textit{Motivation} comprises both conscious and unconscious cognitive processes that guide and stimulate behavior. \textit{Ability} refers to an individual’s psychological and physical ability to engage in a particular activity. \textit{Trigger} is a cue or a call to a particular activity. In the FBM, a \textit{trigger} represents a tangible event that, under the appropriate circumstances, prompts an individual to change their behavior.        

        \item \emph{\textbf{COM-B Model}}~\cite{michie2011behaviour}, standing for \textit{Capability (C)}, \textit{Opportunity (O)}, and \textit{Motivation (M)} is a behavior change model that identifies these three key factors as influential in modifying behavior (B). 
        Although \textit{motivation} and \textit{capability} align with the meanings of \textit{motivation} and \textit{ability} in FBM, COM-B introduces an additional element called \textit{opportunity}. \textit{Opportunity} encompasses external factors that enable or hinder the performance of a behavior. 

        \item \emph{\textbf{Theoretical Domains Framework}}~\cite{michie2005making, atkins2017guide} or TDF identifies 14 empirically verified domains that contain different factors which affect behavior change. TDF \cite{atkins2017guide} consists of 84 factors organized into these 14 domains. The domains include \textit{knowledge}, \textit{skill}, \textit{social role and identity}, \textit{benefits about capability}, \textit{optimism}, \textit{belief about consequence}, \textit{reinforcement}, \textit{intentions}, \textit{goals}, \textit{memory/attention/decision process}, \textit{environmental context and resources}, \textit{social influence}, \textit{emotion} and \textit{behavior regulation}. TDF extends its scope to \remove{explicitly} focus on external social and environmental factors, \remove{thus} providing \remove{researchers with} a more fine-grained \remove{and more comprehensive} framework for identifying factors affecting behavior change\remove{, including both barriers and facilitators that may influence behaviors and target behavior changes}.

    \end{enumerate}

We found that the domains discussed in TDF closely relate to the success of data science. However, the TDF, although good at identifying fine-grained factors affecting behavior change, is difficult to operationalize compared to the COM-B model, and thus has seen fewer direct applications\remove{ than the COM-B model}~\cite{pinder2018digital}. The conciseness and usefulness of the COM-B model\remove{, on the other hand,} is \remove{further} corroborated by the fact that most prior theories~\cite{fishbein2001factors, michie2005making, cane2012validation, atkins2017guide} including TDF\remove{ itself}, ultimately break down into components of the COM-B model. Finally, while COM-B \remove{gives us a good} balances \remove{between} specificity and generalizability, the Fogg Behavior Model~\cite{fogg} is one of the first \remove{established} theories to \remove{begin} consider\remove{ing} Behavior Change Techniques, discussed next.
\remove{It gives us a generative framework to operationalize the COM-B model. Consequently, FBM has been used significantly in prior implementations of Digital Behavior Change Intervention (DBCI) systems~\cite{oinas-kukkonen2009persuasive, pinder2018digital}.} In summary, we \remove{chose to review} \revise{refer to} these three theories of factors affecting behavior change because of the complementary balance they provide in being concise (COM-B), specific (TDF), and operationalizable (FBM).

In~\autoref{fig:examples}, the target \remove{technically satisfactory} behavior for the smoking cessation example could use an increase in the \textit{opportunity} as per COM-B and TDF, to avoid lung cancer by using risk-free cigarettes, thus leading a healthier life. As per FBM, risk-free cigarettes provide a \textit{Trigger} to bring about the change. Under TDF, such a behavior change intervention falls under the \textit{environmental context and resources category} as it alters the resources available at hand. 
\revise{Note that while this solution can reduce toxicity, it does not address the underlying
addiction. We discuss this further in Section~\ref{sec:behaviorally_responsible}.}

On the other hand for the data science domain, the target \remove{responsible} behavior for the machine learning example calls for an increase in \textit{motivation} as per FBM and COM-B to consider and empirically verify the greater impacts of the decisions made from their deployed ML models. As per TDF, this falls under changing the \textit{decision processes} of the data scientist to include this verification step in their workflow. The target \remove{responsible} behavior for the visual data analysis example could be achieved by providing a \textit{trigger} (as per FBM) to the visualization designer to incorporate an evaluation step in the workflow before publishing the visualization. This provides an \textit{opportunity} (as per COM-B) to the designer to verify if their visualizations conform to the \textit{social norms} (as per TDF) of creating visualizations and are therefore effective in conveying the message. 

\subsection{Behavior Change Techniques (BCT)} 
\label{sec:bct}

Behavior change techniques
\remove{are ways of} put\remove{ting} the aforementioned factors of behavior change to work, \ie implementing the \remove{triggers} \revise{interventions} which can bring about behavior change. 
Michie~\etal~\cite{michie2011behaviour} provide a coarse categorization of these techniques as \textit{intervention functions} (IF). However, the most detailed taxonomy in this regard --- the Behavior Change Techniques Taxonomy (BCTTv1), was created by Abraham \& Michie~\etal~\cite{abraham2008taxonomy, michie2013behavior}, which lists 93 such techniques clustered into 16 categories. We use the BCTTv1 taxonomy for its descriptive power and Michie~\etal's~\cite{michie2011behaviour} categorisation for its conciseness, when designing interventions, as described in our illustrative examples in \autoref{sec:interventions}. 

In~\autoref{fig:examples}, for achieving \remove{technically satisfactory} \revise{ the target} behavior, one could \textit{restructure the environment} by providing access to risk-free cigarettes in the smoking cessation example. 
In the data science domain, to achieve the target \remove{behaviorally responsible} behaviors in the machine learning example, organizations could \textit{educate/train} (as per intervention functions) their data scientists to identify possible negative impacts of the decisions of their deployed models on the target groups. This \textit{shaping of their knowledge} (as per BCTTv1) can induce a change in their workflows to include empirical verification of downstream consequences of their model decisions. In terms of the visual data analysis example, the target \remove{behaviorally responsible} behaviors of evaluating visualizations can be achieved through reminding designers to \textit{compare} (as per BCTTv1) their visualizations to the commonly accepted visualization design norms or with visualization design \textit{models} (as per intervention functions) so that viewers do not have difficulties in understanding the conveyed message with the appropriate data context.

\subsection{Mechanisms of Action (MoA)}
\label{sec:moa}
Mechanisms of Action (MoA) represent the processes through which a BCT affects behavior. In other words, it explains \textit{how} a factor of behavior change influences a certain technique to bring about the target change. Carey~\etal~\cite{carey2018behavior} identified 26 different mechanisms of action and linked the behavior change techniques from the BCTTv1 taxonomy to these mechanisms (\eg prompting or giving cues to the subject works by leveraging the \textit{attention} and \textit{}{behavioral cueing} mechanisms of human cognition, both of which affect the \textit{capability} of the subject). We refer back to these Mechanisms of Action to understand the most effective means of designing interventions (\autoref{sec:interventions}) with a theoretical grounding, thereby maximizing impact. 

In~\autoref{fig:examples}, the behavior change techniques of restructuring the environment for the target \remove{technically satisfactory} behavior in the smoking cessation example works through a change in the \textit{environmental context} of the individuals. 
In the data science context (\eg Jupyter Notebook), training the data scientists about the potential ill-effects of their model decisions on the target groups helps in changing the \textit{attitudes towards their behaviors}. One potential behavior change technique -- social comparison in the visual data analysis example works by influencing the visualization designers to adhere to socially accepted \textit{subjective norms} of visualization design while incorporating the data context.


%% file: sections/3.ds.tex
\section{Responsible Data Science}
\label{sec:background_ds}

As a precursor to \remove{establishing} \revise{translating}
behavior change theories \remove{in} \revise{to}
responsible data science, \remove{it is imperative first to} \revise{we must} identify what constitutes responsible data science. \revise{\textbf{Responsible Data Science} includes efforts that address both technical and societal issues. We operationally adopt the definition of responsible data science from Cheng et al\cite{cheng2021socially} which says the objective of Responsible Data Science is to address the social expectations of
generating shared value – enhancing both data science models’ ability and benefits to society.
This definition aligns with existing research examining "Ethical AI" and related topics\cite{mehrabi2021survey,thiebes2021trustworthy,carroll1991pyramid, wang2023designing}. 
} In ~\autoref{sec:outcomes_agents}, we \revise{characterize
responsible data science as a function of \textit{agents} and \textit{outcomes}, with a (typically implicit) role of behavior which
can influence the outcomes.\remove{scaffold this discussion by separating  \textit{agents} and \textit{outcomes} of responsible data science.}} While we briefly review technically satisfactory practices in data science in \autoref{sec:technical_satisfactory}, our primary focus of this paper, elaborated in \autoref{sec:behaviorally_responsible}, is on the aspect of behavioral responsibility.

\subsection{Characterizing Agents and Outcomes of Responsible Data Science}
\label{sec:outcomes_agents}

To scaffold our discussion of responsible data science, we find it useful to characterize it into two dimensions (as shown in ~\autoref{fig:teaser}): \textit{agents} and \textit{desired outcomes}. 
The first dimension, agent, can be \textit{technical} or \textit{human}. 
Technical agents represent systems or techniques used in data science that have the potential to influence the rigor of data science practice through technical indicators, algorithms, systems, and toolkits that are incorporated into the data science project. Human agents, on the other hand, represent behavioral actions that affect the rigor of data science practice. 
\revise{We choose this terminology to emphasize the proactive role of agents within data science and AI. However, our framework is not limited to only AI/data science. It can also be extended into other automated interventions designed with different application scenarios.}

The second dimension, the desired outcome, indicates the extent of attention to care and responsibility paid in the data science practice. 
We categorize desired outcomes loosely as \remove{\sout{either}} \textit{satisfactory} \remove{or} \revise{and} \textit{responsible}. \revise{These outcomes are not mutually exclusive and can overlap.}
Satisfactory outcomes \remove{mean} \revise{focus on \textbf{maximizing benefits} by} following the established best practices without much regard to ethics\revise{; \eg a loan approval model that maximizes the profit of banks but treats applicants who come from different genders unfairly. \remove{While responsible outcomes go above or beyond the current norms of best practices to minimize harm, and actively benefit society.} Responsible outcomes, on the other hand, aim to \textbf{minimize harm} and actively benefit society, incorporating ethical considerations throughout the data science process, \eg a face recognition model that works well for humans from different ethnic groups.} \revise{Moreover, being responsible itself can be seen as an \textbf{attitude} within the data science process, guiding actions and decisions with the intent of delivering responsible results. A responsible data science practice can, and should, encompass both technically satisfactory and behaviorally responsible actions.}

\revise{Among the four combinations of these dimensions shown in Figure 1, we highlight the complementary importance of "Technically Satisfactory" and "Behaviorally Responsible" practices in 
in the frame of responsible data science. "Technically Satisfactory" practices (Figure 1A, green cell) have traditionally been the focus of data science practitioners to ensure that technical aspects of model development are sound, using appropriate tools, models, and metrics. However, they often lack consideration of ethical implications. In contrast, "Behaviorally Responsible" practices (Figure 1D, red cell) emphasize the ethical responsibilities of data scientists
and the broader societal impacts of their actions. This focus on human behavior addresses the root causes of biases and ethical issues that technical solutions alone cannot resolve.} 
\remove{We note that \textit{technically satisfactory} practices (\autoref{fig:teaser}A, green cell) have traditionally been the focus of data science practitioners. 
It involves among many examples, using tools that help us quantitatively evaluate and optimize the model.} in more detail \revise{in Sections \ref{sec:2.2} and \ref{sec:2.3}} and connect them to the examples in~\autoref{fig:examples}.

\subsection{Technically Satisfactory Practices for Responsible Data Science}
\label{sec:2.2}
\label{sec:technical_satisfactory}

Every step in the data science pipeline presents opportunities for decisions that can significantly influence outcomes. \remove{A notable resource in this regard is a comprehensive survey on bias and fairness in machine learning conducted by Mehrabi~\etal\cite{mehrabi2021survey}.} In this subsection, we delve into the specifics of technically satisfactory practices, elucidating key aspects that demand adherence to best practices based on findings from \revise{a comprehensive survey on bias and fairness in machine learning \cite{mehrabi2021survey}}.

\begin{enumerate}
    \item \textbf{Applying appropriate statistical tests:}
    After a research hypothesis is formulated, a suitable statistical test must be used to verify it. However, since domain experts may not be well-versed in statistics, the selection of appropriate statistical tests (\eg one-way or two-way ANOVA) and parameters like significance level must be carefully considered\cite{binnig2017toward,tea}. 
    \remove{Binnig~\etal~\cite{binnig2017toward} provide this kind of statistical guidance to the analyst in their QUDE system. Jun~\etal~\cite{tea} compile high-level specifications into a constraint satisfaction problem that determines the set of valid statistical tests
    . }
    \item \textbf{Applying proper data science models:} The choice of model can significantly impact the quality of results and the ability to make meaningful predictions or decisions~\cite{ding2018model,khaledian2020selecting}. \remove{It begins with a deep understanding of the problem at hand and the characteristics of the data. } Depending on the nature of the data, different models may be more appropriate.
    \remove{, whether it is a decision tree for interpretability, a deep neural network for complex patterns, or a linear regression model for simple relationships}
     Moreover, model selection should consider factors such as scalability, computational resources, and interpretability.
    \remove{, depending on project requirements}
     Regularization techniques and hyperparameter tuning further refine model performance. In some cases, ensemble methods or domain-specific models may be preferred. 
     \remove{The process of choosing the right model should be driven by a comprehensive analysis of the problem, data, and project objectives, ensuring that the selected model is not only technically sound but also best suited to deliver actionable insights or predictions}

    \item \textbf{Applying suitable evaluation metrics:} Applying appropriate evaluation metrics is a pivotal aspect of ensuring a technically sound data science project. It is crucial to align the choice of evaluation metrics with the project's specific objectives \cite{zhou2021evaluating}. Depending on whether the task involves classification, regression, or clustering, different metrics such as accuracy, precision, recall, F1-score, or Mean Absolute Error (MAE) should be carefully considered. Additionally, the presence of imbalanced data or unique business considerations may warrant the use of specialized metrics. Domain knowledge and collaboration with subject matter experts can further guide the selection of metrics that best reflect the real-world impact of the data science solution. 
    
    \item \textbf{Visualizing or communicating results:}
    Correll~\cite{correll2019ethical} calls for communicating the results of data analysis sessions with consideration for the data context and uncertainties, especially when using the medium of visualizations, which can abstract or trivialize the context provided by the data. One example is the proposed use of fuzzy gradient plots instead of well-defined bar charts to better convey the uncertainty in the data~\cite{correll2014error}.
    
\end{enumerate}

Row 2 in~\autoref{fig:examples} illustrates technical, but ethically blind practices\remove{ for each example domain context}.
For the smoking cessation problem, the example of using specialized cigarettes that prevent the risk of lung cancer makes use of technological advancements to achieve the desired satisfactory outcome. 
Extending the analogy to the data science domain, we could use multiple model accuracy metrics to gauge model performance (point 3 in the aforementioned bullet list). In the visual data analysis example, using appropriate empirically verified visual encodings for designing visualizations leads us to the satisfactory outcome of designing good visualizations (point 4 in the aforementioned bullet list).

\subsection{Behaviorally Responsible Practices in Data Science}
\label{sec:2.3}

\label{sec:behaviorally_responsible}

\remove{Complementary to the technically satisfactory practices in data science, there are}
\revise{Several behaviorally responsible approaches complement existing technically satisfactory practices} \remove{that can provide insight into the concept of behavioral responsibility of data scientists,} such \remove{as the approach by} Aragon \etal's\remove{\cite{aragon2022human} who proposed} ethical principles in Human-Centered Data Science~\revise{\cite{aragon2022human}}, Heise \etal's\remove{\cite{heise2019internet} who curated} primary ethical norms in computational research~\revise{\cite{heise2019internet}}, and Zegura \etal's\remove{\cite{zegura2018care} who called for taking} \revise{calls for} care and social good \remove{into consideration} when \remove{conducting} \revise{practicing} data science\remove{ practices}. Among \remove{the prominent} \revise{these} literature\remove{ we reviewed}, we \remove{primarily draw upon the} \revise{emphasize} insights from \textit{Human-Centered Data Science}\cite{aragon2022human} and the concept of \textit{Care and the practice of data science for social good}\cite{zegura2018care}. 
These \remove{two} approaches \remove{offer complementary perspectives because they} \revise{complement one another by} emphasiz\revise{ing} the importance of responsible motivation when \remove{conducting} \revise{practicing} data science \remove{practices} and characterize actionable solutions to facilitate ethical practices in data science. 
While \textit{Human-Centered Data Science} \remove{primarily} emphasizes high-level \remove{principles and} guidelines for practicing data science with care and rigor, \textit{Care and the practice of data science for social good} \remove{delves into the details of} \revise{delineates} responsible practices at each stage of the data science pipeline, such as problem understanding and data preparation\remove{. 
We next}\revise{, which we} describe \remove{these two approaches} in greater detail.

First, \textit{Human-Centered Data Science}\cite{aragon2022human} provides a foundational resource that facilitates a systematic approach to contemplating behaviorally responsible practices within the realm of data science. The book offers a comprehensive set of ethical guidelines that encourage a nuanced consideration of data science projects, \eg ethics on defining the data science problem and ethical principles of training, validating, and testing data science models. The authors emphasize the key characteristic of responsible data science: ``Our goal here is to make you aware that thinking critically and caring about your process and how it affects your results, as well as the people whose behavior is represented in your dataset, is needed every step of the way'' \cite{aragon2022human}. The applicability of these ethical guidelines is notably well-suited to a wide range of situations where humans are, or ought to be, involved, such as loan approval and criminal recidivism predictions. 

Second, the authors of \textit{Care and the practice of data science for social good}\cite{zegura2018care} argued responsible practices are informed by a thoughtful examination of  \textit{how} research is done and in what \textit{context} it is done. It argues responsible data science relies on an ethics approach rooted in practicality: ethics involves not only adhering to formal rules or their definitions but also observing actual behaviors. Ethics shouldn't be treated as a goal to optimize or ``manipulate.''

We assert that ethics requires a continuous process of reflection---considering potential risks, benefits, and \remove{possible} harms. Yet, \remove{ethics goes beyond reflection; being} thoughtful\revise{ness} alone does not \remove{necessarily} prevent harm. \remove{Data scientists might need to collaborate with relevant parties, especially those whose data is utilized in analyses. At each stage of the data science pipeline, new ethical questions and challenges can emerge, demanding attention and resolution.} The imperative to embrace responsible behaviors in data science emerges from the recognition that a
\remove{singular, fixed, and} 
standardized checklist is often insufficient across diverse scenarios encountered in the field \cite{aragon2022human}. Instead, behaviorally responsible data science practices demand that practitioners proactively \revise{cultivate and dynamically respond to their}
\remove{apply thorough attention and resolution to dynamically consider the nature of the} specific data science problem and context. Thus, a reflexive and adaptable stance is essential, acknowledging that the ethical considerations surrounding each data science project are nuanced and distinct. 
This diversity emphasizes the pivotal role of "\textbf{Care Ethics}\cite{sanchez2018big, aragon2022human, zegura2018care}", a key concept to both \textit{Human-Centered Data Science} and the concept of \textit{Care and the practice of data science for social good} as a foundational principle guiding behaviorally responsible data science. 

Care Ethics encourages data practitioners to approach their work with a deep sense of empathy and conscientiousness\cite{zegura2018care, meng2019collaborative}. \revise{For example, Zegura et al\cite{zegura2018care} proposed an orientation to a caring mindset in the practice of data science that facilitates social good; Meng et al\cite{meng2019collaborative} highlighted the importance of applying the ethics of care in democracy within collaborative data work.} \remove{Care Ethics serves as a catalyst for thoughtful consideration of the potential risks and harms associated with decisions
and actions taken
during the data science process.
By embracing this principle, data scientists adopt a mindset that prioritizes minimizing harm and maximizing benefit to all stakeholders involved.}
This approach prompts practitioners to reflect on how their choices would impact individuals, communities, and society at large. The notion of Care Ethics introduces a transformative shift in perspective\cite{boone2023data}. Encouraging data scientists to envisage their data science projects as endeavors involving their own family and loved ones cultivates a heightened sense of responsibility.
\remove{This shift compels practitioners to go beyond technical and methodological considerations, delving into the profound societal and ethical implications of their work.}
Promoting Care Ethics not only enhances the behavioral responsibility of data science but also infuses the decision-making process with an intrinsic sense of accountability. 

Inspired by the principles contained within care ethics\cite{sanchez2018big, aragon2022human, zegura2018care}, we review some actionable activities that align with behaviorally responsible data science. 
These practices are not exhaustive and should not be viewed as a checklist -- instead, these serve only as inspirational examples to further ground the concept of behavioral responsibility in data science.

\begin{enumerate}

    \item \textbf{Comprehensive problem understanding: } 
    Understanding the \revise{influence of bias in} data science problem\revise{s} \remove{in an unbiased way} is fundamental to behavioral responsibility in data science. \remove{It starts with a clear recognition that preconceived notions or biases should not shape problem formulation\cite{zegura2018care}. Data scientists must approach the problem with an open mind, free from personal or organizational biases that could skew the analysis or results \cite{van2016data}.} \revise{Data scientists should be aware of their own biases and how these biases affect the way they formulate the problem\cite{van2016data}. To counteract these biases,} it is essential to involve diverse perspectives and stakeholders to \remove{ensure a comprehensive} \revise{develop a nuanced} understanding of the problem\remove{'s nuances} and potential impacts on different groups. Beyond that, attention should be paid to examining historical data, and considering the historical context of the problem as it could reveal biases or systemic inequalities that need to be addressed. \remove{This process also allows our partners to elaborate on what the community cares about and how the community envisions putting its care into practice through data.}

    \item \textbf{Collecting unbiased data:} \remove{Attention should be paid to unbalanced datasets because} \revise{Imbalanced datasets} can lead to biased models that perform poorly on minority classes. Data science practitioners may consider \remove{solving this issue by} gathering more data for \remove{the} minority classes, oversampling minority classes, or \remove{increasing the weight of} \revise{re-weighting} minority class\revise{es} to \remove{deal with} \revise{address} the issue \cite{viloria2020unbalanced}.
    Systems like Trifacta~\cite{trifacta} enable dataset anomaly detection and quality assessment using quality rules such as data integrity constraints. Apart from that, consideration \remove{also needs to} \revise{must} be given to data points that do not \remove{even} \revise{yet} exist in the \remove{collected} data\remove{set}~\cite{gitelman2013raw}, which \revise{may result in} \remove{can potentially give us} a biased starting point. 

    \item \textbf{Careful data preparation: } \remove{Another pivotal step in the data science journey is} \revise{D}ata preparation\remove{, which} holds immense significance for behavioral responsibility in the field. \remove{The} \revise{This} process \remove{of data preparation} involves \remove{meticulously} cleaning and wrangling datasets\remove{,} \revise{to ensure} that the data is accurate, complete, and free from bias \cite{brownlee2020data,van2016data}.
    \revise{In addition to technically responsible techniques} such as handling missing values, outlier detection and treatment, and \revise{thoughtful feature engineering,} \remove{addressing data imbalances are crucial in creating a reliable foundation for subsequent analyses. In terms of caring data science, }
    behaviorally responsible data preparation extends to the responsible handling of sensitive information, anonymizing data when necessary, and safeguarding privacy to uphold behaviorally responsible standards. \remove{Transformation techniques, such as feature engineering, enable the addition of meaningful attributes that contribute to a deeper understanding of the problem.}

    \item \textbf{Identifying biased interactions with data:}
    Identifying when bias may occur during analysis or interpretation, especially during interactive data analysis where the analyst may selectively look at certain data points while neglecting others (even though inadvertently)~\cite{kahneman2011thinking} is also a crucial step. Wall~\etal~\cite{wall2017warning,wall2021left} propose an approach of computing and visualizing bias in user interactions during visual analysis. \remove{Zgraggen~\etal~\cite{zgraggen2018investigating} propose confirmatory strategies to account for repeated testing or the multiple comparisons problem, to avoid \textit{p-hacking}, during exploratory visual analysis.}
    
    \item 
    \remove{\textbf{Provenance tracking for accountability:} It is easy to get side-tracked due to more interesting but illusory patterns in data, and enter into a rabbit hole. To avoid such situations, it becomes necessary for the analysts to be aware of the individual steps during analysis { as well as their high-level approach.} To facilitate this, visual analysis systems with provenance tracking have been created~\cite{callahan2006vistrails,feng2017hindsight,wood2019design} to help keep analysts grounded.} 
    \revise{\textbf{External Reviews for Accountability:} Accountability in data science should extend beyond technical reviews to include assessments by peers and stakeholders who will be impacted by the model. This involves treating the review process not just as a technical code review but also as a review of ethical practices and implications. 
    These reviewers can identify potential harm and unintended consequences that may not be evident to the technical team. One way to support this type of review is to support provenance tracking~\cite{callahan2006vistrails,feng2017hindsight,wood2019design}, so that data scientists may be held accountable not only to the outcomes of their models, but their process as well.} 
    
    \item \textbf{Streamlining pipelines with checklists}: 
    At the commercial level where stakes are typically high, checklists have been created to draw developers' attention to the entire pipeline specifically in machine learning-based data science~\cite{jobin2019global, cramer2019translation, madaio2020co-designing}. Notable tasks within these checklists among many others, include ensuring fairness and privacy during data collection\remove{(\eg Is the population subject to inferences drawn from the collected data, correctly represented in the data? Do we have appropriate consent to gather the data?)}, transparency during analysis\remove{ (\eg Can we reproduce the steps performed during this analysis?)}, and interpretability during inference\remove{ (\eg Which factor was given more weight in a certain decision?)}.
\end{enumerate}

\revise{Revisiting the smoking cessation example}
\remove{Connecting behaviorally responsible practices to the examples} in~\autoref{fig:examples}\remove{, for the smoking cessation example}, a smok\revise{er} \remove{addict} may act responsibly \remove{and} \revise{by increasing} his awareness \remove{about} \revise{of} the health consequences of smoking to \remove{get rid of} \revise{fight} his addiction. 
Extending the idea to the machine learning example \remove{in the data science domain}, a data scientist could act responsibly by evaluating the decisions of an ML model \remove{empirically, thereby becoming capable of anticipating any} \revise{for potential} downstream consequences (point 1 in the aforementioned \remove{bullet} list). In the visual data analysis example, \remove{one example of} responsible behavior could be to \revise{involve external} evaluat\revise{ion of} the visualizations \remove{designed for dissemination to check if they are able to convey the message clearly (point 6 in the aforementioned bullet list)} \revise{
to check if they convey information in a non-misleading way (point 5 above)}.

%% file: sections/operationalize.tex
\section{Operationalizing Behavior Change
Theories for Responsible Data Science}
\label{sec:operationalize}

In the preceding sections, we \revise{described} the rich landscape of behavior change theories\remove{, exploring their nuances and applications in} \revise{for} data science. As we transition from a theoretical understanding to practical applications, it is essential to reflect on how these theories can be operationalized to design interventions in real-world data science scenarios. This critical step involves not only identifying and addressing specific behaviors within the context of data science but also decomposing the design process of behavior change interventions in the \remove{digital} context of data science environment. In this section, we aim to bridge this gap by offering a guide to translating theoretical insights into actionable steps.

In the previous section, behavior change theories were introduced chronologically based on the date of the publication of theories (e.g., factors of behavior change~\cite{fogg,michie2011behaviour,michie2005making, atkins2017guide}, behavior change techniques~\cite{abraham2008taxonomy, michie2013behavior}, and most recently work towards understanding mechanisms of action~\cite{carey2018behavior}). In this section, we alter this order to align with how we think about operationalizing these theories towards the development of interventions for responsible data science.

\begin{enumerate}
    \item \textbf{Identify problematic and target behaviors:} It is crucial to pinpoint both problematic behaviors that might impede responsible data science practices and target behaviors that should be encouraged to replace them \revise(see \autoref{sec:2.2} and~\autoref{sec:2.3} \remove{mention some} \revise{for} examples\remove{ of target behaviors}\revise{)}. This requires analysis of current methodologies and workflows\remove{, emphasizing behaviors that must change to align with ethical and responsible data science}. For instance, overlooking biases in data or algorithms can be considered a problematic behavior in the context of responsible data science, whereas actively seeking diverse data sources might be a target responsible behavior to encourage.
 
    \item \textbf{Identify factors affecting problematic behaviors:} Building on the theories outlined in \autoref{sec:3.1}, we need to identify various factors (\eg \textit{capability, opportunity, and motivation}\remove{ introduced in the COM-B model}~\cite{michie2011behaviour}) that might influence the problematic and target behaviors identified in the previous step. We then need to assess whether digital interventions are appropriate given the factors involved.
    \remove{This involves exploring cognitive biases, structural and systemic issues in the field, and the influence of organizational culture on data science practices.} 
    For instance, insufficient training can perpetuate undesirable practices in data cleaning, which might be rectified through interventions aimed at enhancing \textit{capability}. \remove{Similarly, interventions designed to improve understanding of the ethical consequences of model deployment could boost \textit{motivation}, thereby encouraging more ethically responsible behavior.} 
    \revise{Similarly, the lack of awareness among data science practitioners regarding the potential social impacts of their models can jeopardize the benefits to affected groups. This gap can be bridged by interventions that enhance their \textit{motivation} to understand the ethical consequences of the models they develop.}
    
    \item \textbf{Understand and employ appropriate Mechanisms of Action:} Once the factors affecting the problematic behavior are identified, the appropriate mechanism of inducing the target behavior needs to be identified and employed, as discussed in~\autoref{sec:moa}. This involves understanding how different strategies leverage \textit{capability, opportunity}, or \textit{motivation} to initiate and sustain behavior change among data science professionals. \remove{For example, educating data scientists about the potential outcomes of their models in the machine learning example (\autoref{fig:examples}) in the environment can promote responsible behaviors through changing \textit{attitudes towards their behaviors} and updating their \textit{beliefs about consequences}~\cite{carey2018behavior}.} 
    \revise{Taking the machine learning scenario as an example (\autoref{fig:examples}), if data scientists lack \textit{motivation} to commit more time to test model outcomes on different influenced groups, this could be bridged the mechanisms related to changing their \textit{attitudes towards their behaviors} and updating their \textit{beliefs about consequences}~\cite{carey2018behavior}.}
    \remove{This foresight involves evaluating the context-specific applicability of each intervention, predicting possible outcomes, and understanding the nuances of their implementation in the data science environment.} This not only helps designers to choose the most appropriate interventions for the digital context but also facilitates them to maximize impact. 
    
    \item \textbf{Envision potential interventions using BCT:} Having identified both the factors affecting problematic behaviors and the underlying mechanism of action, we can now envision potential interventions in the \remove{digital} data science context by referring to the behavior change techniques~\cite{abraham2008taxonomy, michie2013behavior} introduced in~\autoref{sec:bct}. These might include training programs, ethical guidelines, and decision-support tools that encourage reflection on the consequences of one’s actions in the data science workflow. For example, to develop interventions in the data science environment that boost \textit{motivation} by strengthening the understanding of ethical implications in the machine learning example (\autoref{fig:examples}), 
    organizations can educate or train their data scientists\revise{, which facilitates their \textit{beliefs about consequences} and \textit{knowledge}}. This \textit{shaping of their knowledge} (as per BCTTv1\cite{michie2013behavior}) could focus on identifying potential negative impacts of their models on target groups, using a variety of illustrative case studies. \\

\end{enumerate}

\revise{Prior work has mapped factors affecting behavior change (FBCs) to specific interventions (BCTs)~\cite{steinmo2015characterising} and specific interventions to their underlying mechanisms of action (MoAs)~\cite{carey2019behavior}. To help designers choose appropriate FBCs, MoAs, and interventions, we provide a supplemental table that merges these mappings.}

%% file: sections/4.interventions_14_sept.tex
\section{Interventions}
\label{sec:interventions}

\remove{Given the responsible behaviors and theories of behavior change in the context of data science in the previous sections,
we operationalize these theories next to envision some example behavior change interventions in data science.}

In this section, we refer back to the two data science contexts in \autoref{fig:examples} to \revise{apply these theories and} discuss potential interventions \remove{and the behavior change theories behind them} for both the desired technically satisfactory and behaviorally responsible practices for the machine learning example in the second column (\autoref{sec:4.1}), and visual data analysis example in the third column (\autoref{sec:4.2}). Note that this is not an exhaustive account of interventions for these two contexts but merely describes some possibilities, grounded in behavior change theory. \revise{We further use this as an opportunity to describe these two examples as \textbf{usage scenarios} to explain how to apply the framework from Section~\ref{sec:operationalize}, and in ~\autoref{sec:internal_reflection}, we provide our own \textbf{internal reflection} on the usage of this framework for envisioning behavior change
interventions for responsible data science.} 

\subsection{Interventions Designed for the Machine Learning Example}
\label{sec:4.1}

\remove{
\paragraph{Context: } Recalling the machine learning example in \autoref{fig:examples}, to the end of ensuring data scientists deliver accurate, truly generalized, unbiased models, interventions can be designed to educate data scientists about possible harmful impacts
of model decisions on the potentially disadvantaged groups in social-good sensitive tasks \eg female applicants who are usually systematically discriminated against in the loan approval prediction algorithms\cite{turner1999mortgage}.
We next describe the behaviors and relevant theories to envision possible solutions, as outlined in~\autoref{sec:operationalize}:}

\remove{\paragraph{Identify problematic and target behaviors:} In this case, the target behavior is to increase data scientists' awareness and stimulate their empathy regarding the effects of their model decisions on socially disadvantaged groups, particularly in tasks sensitive to social good.}

\remove{\paragraph{Identify factors affecting problematic behaviors:} One possible explanation we will focus on in this example is that appropriate \textit{motivation}
may be lacking. Educating data scientists about possible harmful impacts of model decisions on the potentially disadvantaged groups in social-good sensitive tasks can increase their \textit{motivation} (FBC)~\cite{michie2005making} to identify any fallouts and take corrective measures accordingly during their data science practices. }

\remove{\paragraph{Understand and employ appropriate Mechanisms of Action:} If \textit{motivation} is lacking, an underlying
mechanism that can explain and influence this is \textit{attitudes towards their behaviors} 
(MoA)~\cite{carey2018behavior}. The perspective of data scientists towards the consequences of their actions can be affected over time, and this mechanism can eventually increase the \textit{motivation} of the data scientists.}

\begin{figure}
    \centering
    \includegraphics[width=0.9\textwidth]{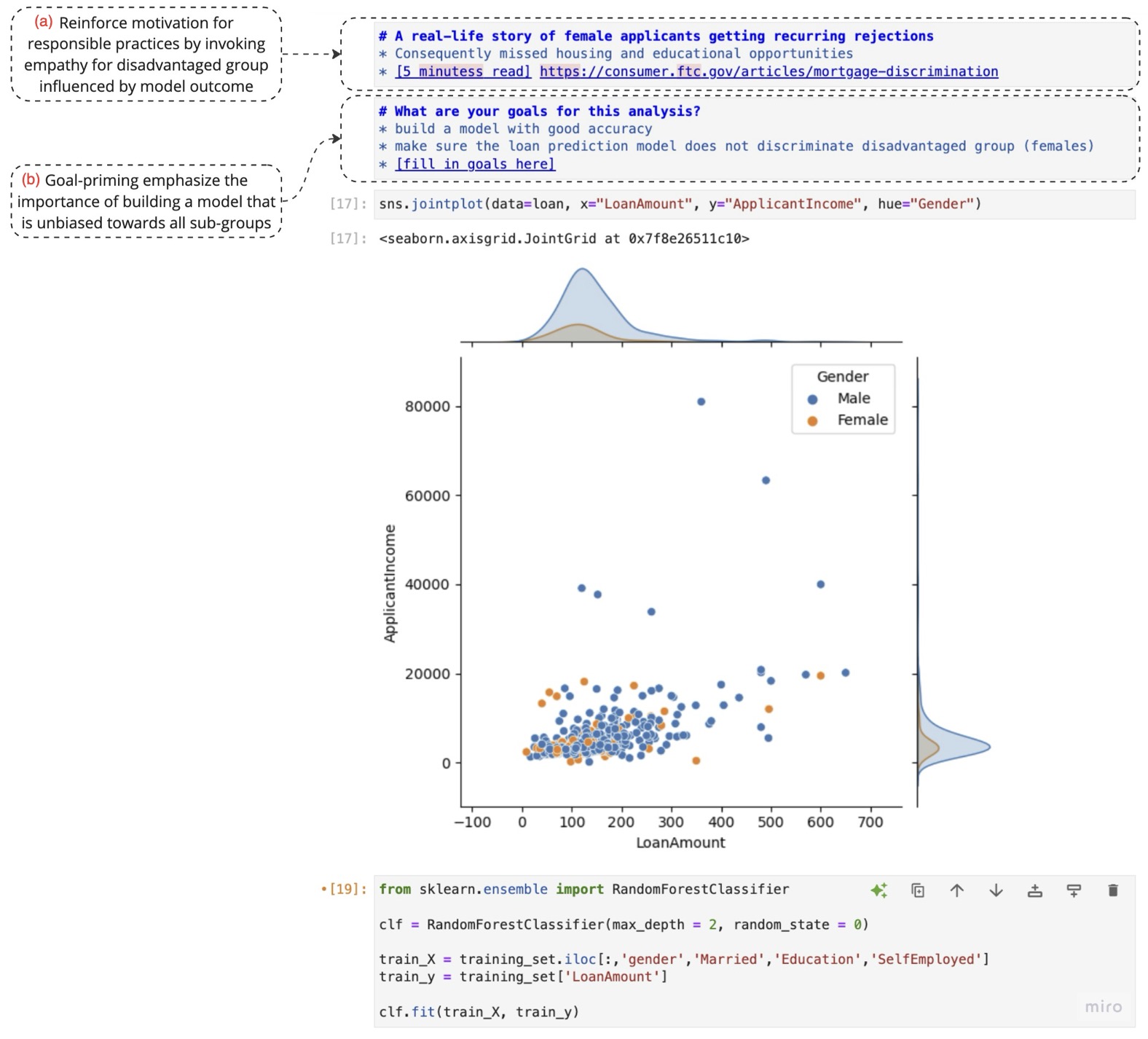}
    \caption{
    As data scientists start analyzing the loan approval dataset within a Jupyter notebook, this intervention (a) reinforces their motivation to practice responsible data science by sharing a real-life story that highlights the potential harm that model outcomes can inflict on disadvantaged groups, aiming to evoke their empathy; (b) follows-up with a goal-priming hint to emphasize the importance of behaving in an unbiased way towards vulnerable sub-groups that are influenced by the model’s outcome. 
    }
    \label{fig:loan}
\end{figure}

\remove{\paragraph{Envision potential interventions using BCT:} Potential interventions could employ behavior change techniques (BCT) of \textit{shaping knowledge} of possible downstream consequences through \textit{antecedents} of previous cases of \textit{social consequences} or inequities in loan approval systems, followed by \textit{setting goals} to address the consequences~\cite{michie2013behavior}. Techniques can be designed to assist data science managers in establishing behaviorally-responsible notebook environments that incorporate built-in tools for providing in-context notes and reminders, thereby facilitating responsible practices. 
Specifically, one potential intervention that could be incorporated in a notebook could involve exposing data scientists to real-life stories of female applicants who encountered recurring rejections from loan approval models, consequently missing out on housing or educational opportunities. This intervention design can reinforce a data scientist's \textit{motivation} for behaviorally responsible practices by invoking their empathy for potentially disadvantaged groups and priming analysts to one particular goal in this socially-conscious data science task. This exposure could take the form of introductory background contexts generated at the beginning of a data science project~(\autoref{fig:loan}.a), followed by a goal-priming hint to verify that the model is behaving in an unbiased way towards vulnerable sub-groups that are influenced by the model's outcome~(\autoref{fig:loan}.b). 
}

\revise{Maggie is a researcher who is designing Jupyter Notebook plugins to help people build more socially responsible models. She is collaborating with a data science team tasked with creating a loan approval model that avoids discriminating against potentially disadvantaged groups, such as female applicants~\cite{turner1999mortgage}. Maggie decides to implement interventions based on our proposed framework.}

\revise{She begins by \textbf{identifying problematic and target behaviors} within the team’s workflow that could hinder the development of a fair model. Maggie is certain that the team has sufficient expertise to tackle the technical challenges of data science models and deliver models with high accuracy. However she is concerned that they may not have enough understanding of how decisions made during the process of data wrangling and model building can have downstream effects, influencing the outcomes for different potentially disadvantaged groups. 
\textbf{Recognizing a lack of motivation as a significant factor}, Maggie wants to 
increase their awareness and empathy regarding the effects of their model decisions on socially disadvantaged groups. 
}

\revise{To achieve this, Maggie thinks that she should \textbf{change the \textit{attitudes towards their behaviors} (MoA)~\cite{carey2018behavior}} and employs this mechanism within her interventions. In the form of real-life stories of individuals, particularly female applicants who have faced repeated rejections from loan approval models, resulting in missed opportunities for housing or education, Maggie uses \textbf{sharing \textit{social and environmental consequences} (BCT)~\cite{michie2013behavior}}. These stories are integrated into the team’s data analysis environment 
\revise{within a notebook cell shown on the top, containing hyperlinks for these stories (\autoref{fig:loan}.a)}, serving as a constant reminder of the real-world impacts of their work.}

\revise{Additionally, she \textbf{identifies \textit{goals} (MoA)~\cite{carey2018behavior}} as another possible underlying mechanism and decides to employ \textbf{\textit{goal priming} (BCT)~\cite{michie2013behavior}} intervention within the data analysis tools. She introduces prompts in the data analysis environment to ensure that data scientists are explicit about their goals throughout the workflow (\autoref{fig:loan}.b).}

\revise{Enhancing the workflow in this way with contextual anecdotes and prompts to elicit development goals ensures that data scientists continuously reflect on the ethical aspects of their work. By envisioning potential interventions using our framework, Maggie ensures that her team is not only technically proficient but also behaviorally responsible.}

\subsection{Interventions Designed for the Visual Data Analytics Example}

\label{sec:4.2}

\begin{figure}
    \centering
    \includegraphics[width=0.75\textwidth]{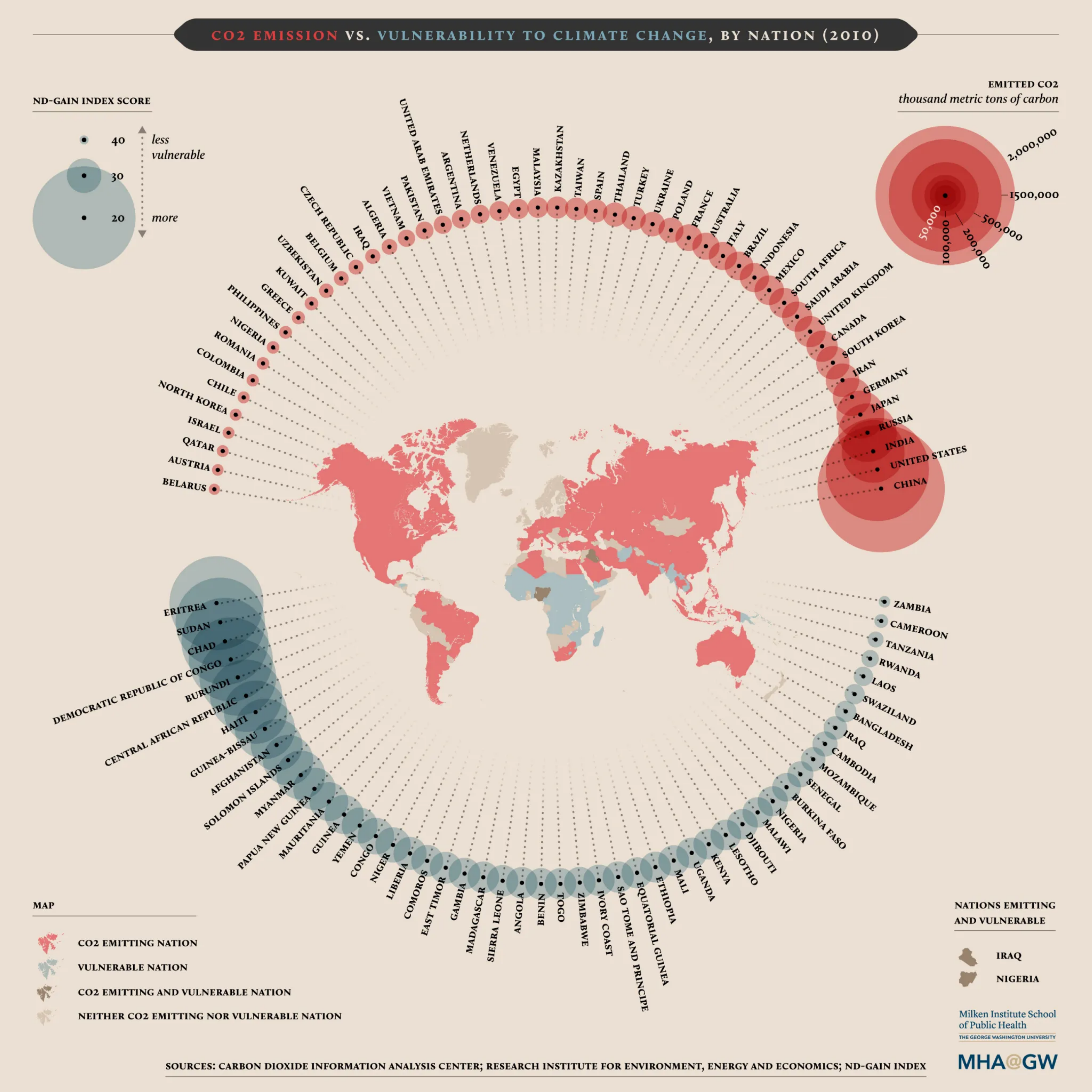}
    \caption{A data visualization showing which nations are major CO2 emitters, and which nations are vulnerable to the effects of these emissions. \revise{In its current state, this visualization might only help global policymakers like the Intergovernmental Panel on Climate Change (IPCC).} By gathering feedback from viewer groups of different backgrounds like politicians, farmers, and students, this visualization could be made more effective by additionally visualizing how each group contributes to these emissions and how they could help alleviate the problem. Credits: https://onlinepublichealth.gwu.edu/resources/climate-change-emissions-data/}
    \label{fig:vda-example}
\end{figure}

\remove{
\paragraph{Context: }In the case of the visual data analytics example in \autoref{fig:examples}, where the objective is to effectively demonstrate the data science outcomes to a diverse audience, the creation of clear and trustworthy visualizations that resist (implicit and explicit) biases poses a significant challenge\cite{cairo2019charts}. This challenge is particularly daunting when the audience hails from various backgrounds (\eg farmers, high school students, politicians) and possesses different levels of domain expertise and sociopolitical perspectives~\cite{peck2019data}. For example, consider~\autoref{fig:vda-example} which estimates the carbon emissions from different countries, and highlights countries that are vulnerable to the effects of these emissions. In its current state, it may only be helpful for policymakers or government officials to identify how they could contribute to solving this problem and may miss the mark for effectively communicating with audiences such as students or farmers, who could contribute to this problem in their own individual capacities. }

\remove{
\paragraph{Identify problematic and desired behaviors:} In this case, the desired behavior is to adeptly convey the data science outcome through data visualizations to audiences with diverse backgrounds.}
    
\remove{\paragraph{Identify factors affecting problematic behaviors:} 
One possible reason why data scientists struggle to communicate their model outcomes to diverse audiences is the lack of adequate \textit{opportunity} (FBC)~\cite{michie2005making}. If data scientists are provided with sufficient \textit{opportunity} to hear needs from diverse groups, they can better understand the primary questions these audiences have about the visualizations and thus create clear and trustworthy visualizations.}

\remove{\paragraph{Understand and employ appropriate Mechanisms of Action:} Mechanisms of \textit{beliefs about consequences} (MoA)~\cite{carey2018behavior} could be employed in this case. These MoAs would introduce data scientists to additional \textit{Opportunity} in the form of feedback from user groups, for improving their visualization designs. }

\remove{\paragraph{Envision potential interventions using BCT:} One possible approach could involve prompting visualization designers to remind them of the target audience, which falls under \textit{feedback and monitoring} (BCT)~\cite{michie2013behavior}. This includes feedback from both the visualization design tool and feedback from the end-users of the visualization. To empower data scientists in this task, interventions could be designed to prompt them to describe the anticipated range of their target audience~\cite{lee-robbins_affective_2022} (\eg policymakers, urban public, university students, rural public, \etc). Then, questions could be generated to match what audience members might ask about the visualization from different perspectives; for example, how different agricultural activities contribute to these emissions might be of interest to the rural community; what steps can students take within their capacity to help alleviate this problem or spread further awareness about it?
To encourage data scientists to answer these questions by interacting with real people, potential interventions could generate customizable annotations that appear upon completion of visualizations along with a shareable link. These annotations nudge data scientists to proactively communicate and gather feedback from stakeholders from diverse backgrounds, capturing their individual perspectives and levels of comprehension regarding the visualizations, and facilitating data scientists in pinpointing potentially confusing or divisive areas in need of improvement. }


\revise{
Dylan is a quality control specialist within a data visualization team. His primary objective is to ensure that his team creates clear and trustworthy visualizations that are non-misleading and can be easily understood by people with various educational/occupational backgrounds. }

\revise{For example, consider ~\autoref{fig:vda-example} which estimates the carbon emissions from different countries, and highlights countries vulnerable to the effects of these emissions. The current version may only be helpful for policymakers to identify how they could address this problem, and may miss the chance to communicate with audiences such as students or farmers, who could address this problem in their own individual capacities. Realizing the scope to maximise impact, Dylan decides to design an intervention tool based on our proposed framework to help the visualization team. }

\revise{While \textbf{identifying problematic and target behaviors} in the visualization workflow, Dylan realizes that although technically adept, the designers in his team might not realize how many different target audiences may encounter their visualizations. He considers this problem in terms of a textbf{lack of motivation}, and decides to use the \textit{social influences} MoA~\cite{michie2013behavior} to help them realise the potential impact on society. Dylan designs interventions to prompt designers to estimate the anticipated range of their target audience~\cite{lee-robbins_affective_2022} (\eg policymakers, urban public, university students, rural public, \etc). He generates possible scenarios to expose designers to diverse questions audience members might ask about the visualization; \eg how different agricultural activities contribute to these emissions might interest rural communities; students might be interested in the steps they can take to help alleviate this problem or increase awareness.}

\revise{Alternatively, Dylan also \textbf{recognizes a lack of opportunity as a factor}. He understands that although the designers are aware of the different target audience groups, not getting feedback from these diverse groups is hindering them from creating more inclusive and effective visualizations.}

\revise{To achieve this, Dylan \textbf{incorporates \textit{feedback processes (MoA)}~\cite{michie2013behavior}}. Using our framework, Dylan figures out that  gathering feedback from different target audience groups, which falls under \textbf{\textit{feedback of behavior} (BCT)~\cite{michie2013behavior}} could be employed. Dylan thus generates customizable annotations that appear upon completion of visualizations along with a shareable link. These annotations nudge the designers to proactively communicate and gather feedback from stakeholders from diverse backgrounds, capturing their individual perspectives and levels of comprehension regarding the visualizations, and facilitating the designers in pinpointing potentially confusing or divisive areas in need of improvement.}

\subsection{\revise{Internal Reflection}}
\label{sec:internal_reflection}

\revise{In this section, we reflect on the usage of our proposed framework for envisioning behavior change interventions for responsible data science. 
We do so by reflecting on some explicit questions.}

\revise{\textbf{Where was the framework the most helpful?}}
\revise{As seen in the previous two subsections, the framework helped Maggie and Dylan enlist multiple possible FBCs, and the possible MoAs and corresponding BCTs to bring about the desired behavior change. Maggie used the framework to identify different MoAs (\textit{attitude towards behavior} and \textit{goals}) to better motivate the data scientists to consider the downstream effects of their models. On the other hand, Dylan found two different FBCs - \textit{motivation} and \textit{opportunity} and accordingly employed BCTs to help visualization designers. The framework thus acted as a comprehensive, though not necessarily exhaustive tool, to generate ideas in a systematic way, without which both Maggie and Dylan could have missed out on potential additional ways to bring about responsible behavior change.
}

\revise{\textbf{Where was the framework the least helpful?}}
\revise{The framework provides a consolidated space of possible approaches for identifying the scope for responsible behavior change, and actionable techniques to bring about the change. However occasionally, the boundaries between the individual FBCs, MoAs, and BCTs that are applicable in a situation are not very clear. For example, \textit{attitude towards behavior} MoA could be employed through BCTs of both \textit{consequences} and \textit{rewards}. The appropriate alternative is evident based on the context in most cases, \eg Maggie used the MoA to inform data scientists about the \textit{consequences}. However, the blurred boundaries or redundancies between some of these terms might cause difficulties for practitioners to use the framework.}

\revise{Further, there is an open-ended nature in the interpretation of a certain situation as lacking a certain FBC. For example, Maggie's intervention of providing \textit{prompts for goal priming} could boost \textit{motivation} of the data scientist to also address downstream consequences of their models. However, it could also be interpreted as providing \textit{opportunity} to the data scientist through \textit{prompts/cues} during model development to inform them about downstream consequences. The latter interpretation assumes that the data scientist is already motivated but lacks the right opportunity to be reminded about responsible behavior.}

\revise{Although these limitations of our framework might create complications in choosing the appropriate BCT, the framework also helps practitioners by making them aware of the possible multiple interpretations of the situation. We thus see this framework as providing a full range of behavior change solutions, while leaving the responsibility to choose the right alternative to the practitioner.}



%% file: sections/5.discussion.tex
\section{Open Research Challenges}
\label{sec:discussion}

In this paper, we introduced a new perspective on responsible data science that elevates the importance of responsible \textit{agents} through the lens of behavior change. Our research complements existing work in guideline and curriculum development by encouraging analysts to adopt more responsible analysis behaviors through their real-time interactions with data science tools. 
Based on this novel perspective,
we identify pressing research challenges moving forward.

\subsection{Challenge 1: Intervening at the Right Time}
\label{sec:disraptor}

When introducing interventions to foster responsible data science practices, it is crucial to strike a balance where interventions are neither absent when assistance is needed nor persistent to the point of causing frustration\remove{ for data scientists}. 
Hence, the timing and appropriateness of interventions are crucial not only for their effectiveness but also for ensuring a positive user experience\remove{ among data scientists}. 
Adopting the concept of \textit{triggers} for behavior change interventions \cite{fogg}, we could conceptualize \textit{when} to initiate an intervention as a \textit{disruptor}.  We present several heuristic approaches to \textit{disrupt} data science practices to initiate an intervention, hoping to inspire potential solutions to this open challenge:

\par{\textbf{Disrupt by Data Science Phase.}} The output at each phase of a data science workflow serves as input to, or otherwise influences, the next phase\remove{ in the pipeline}. 
A seemingly \remove{minor or} benign negligence of technically satisfactory or behaviorally responsible practices in one phase can significantly impact downstream deliverables through the propagation of inaccuracies and biases. Thus, the beginning or end of a phase in the data science pipeline may be an apt time to disrupt the user's process to intervene. \remove{\eg remind the data scientist to check for class imbalance scores after the data procurement phase}

\par{\textbf{Disrupt by Algorithmic Performance.}} 
Interventions could also be designed to disrupt a data scientist's practices based on active monitoring of metrics. In a fairness-aware data science project, \eg crime recidivism analysis, algorithmic bias is treated as an important evaluation metric.  
One disruptor, in this case, could be to continuously monitor fairness metrics that have been identified as critical for the task to identify important dips to the metrics of interest when data is transformed, model parameters are tuned, etc.

\par{\textbf{Disrupt by Minimal Timeline.}} 
Another possible disruptor could be related to minimum expectations for time spent on some tasks. 
The underlying assumptions are that (1) some tasks must be completed, \eg running fairness checks on data prior to model building, and (2) there is an expected amount of time to complete various tasks, which when below a threshold, may be indicative of negligence to fully understand the data or model. 
\remove{In a data science project, the efforts and time commitment to each task are different depending on the nature and complexity of the problem domain and the stage in the pipeline (\eg data cleaning, feature selection, hyperparameter tuning, etc. may require different time commitments). Hence, a reasonable starting point for ensuring the behavioral responsibility of data scientists could be to disrupt an analyst's process if they complete a task significantly earlier than some expected completion time.}
For instance, if an analyst does not dedicate time to exploring the data prior to creating a model from it, it could lead to unknown problems downstream. 
\remove{An intervention could be introduced to disrupt the process to prompt the analyst's confirmation and reflection that the data is sound before progressing.}

\par{\textbf{Disrupt by Third-party Review. }} 
Interventions could also involve an external ethics review board or third-party auditors to conduct periodic assessments of the data science project to provide an unbiased evaluation of responsible data science practices. Furthermore, interventions can be based on feedback from stakeholders, end-users, or affected communities to address any ethical issues or concerns that arise during the project's lifecycle.

\par{\textbf{Disrupt by Programming Specification.}} 
Disruptors can also be identified through established issues in technical standards. For example, a normative coding style is beneficial to avoid ambiguity for implementer reading, collaboration purposes, and future model maintenance. Non-normative coding and variable naming habits may impede collaborators from easily verifying the code which can discourage or hinder future checks on responsible practices. 
\remove{Thus disruptors could be designed to identify violations of established programming patterns} \revise{Thus, disruptors could identify violations of established programming patterns.}

\subsection{Challenge 2: Facilitating Lasting Behavior Change Through In-The-Moment Interventions}
\label{sec:discussion-habit}

In this paper, we focus primarily on settings and examples for in-the-moment interventions that potentially result in short-term positive behavior changes during data analysis. However, it is unclear \textit{if} and \textit{how} these interventions will fundamentally change the long-term practices of analysts~\cite{pinder2018digital}. We view in-the-moment interventions as a subset of behavior change techniques (BCT) for facilitating rigorous data science. The superset also includes interventions for long-term behavior change or habit formation, \eg provide learning materials like enrolling in a course to \bct{learn new skills}, checklist generation (\bct{Planning} and \bct{Repetition}), \bct{setting goals} and \bct{tracking progress}~\cite{lally2013promoting}. Alternative theories and applications of behavior change interventions emphasize settings intended to encourage longer-term habit formation~\cite{fjeldsoe2009behavior, steinmetz2016effective}.
Accordingly, we observe \revise{at least three} major challenges in establishing lasting behavior changes in analysts.

\textbf{Ensuring smooth hand-offs} between in-the-moment interventions for short-term behavior change, and interventions for long-term behavior change will hopefully \remove{ensure that} \revise{expose} analysts \remove{are exposed} to a range of experiences to reinforce rigorous data science practices. Examples include pointing the analyst to relevant online tutorials or courses on statistical testing (long-term) \textit{after} detecting improper statistical testing being performed
and recommending more appropriate tests (short-term).

\textbf{Accounting for the evolution (or devolution) of the analyst.} The analyst's practice of data science may change over time, which may influence both the efficacy of interventions and disruptors. For example, disrupting the flow of \remove{an} \revise{a confident} analyst \remove{who is confident in completing} \revise{during} a well-defined task may hinder rather than accelerate their work~\cite{horvitz1999}. How then do we interrupt an analyst who is initially receptive to interventions, but starts ignoring them later for an unknown reason? \remove{This also points to our next challenge of defining successful versus unsuccessful interventions and disruptors.}



\revise{\textbf{Tackling Long-term Bias. }Another critical aspect of designing long-term interventions is considering the potential long-standing biases that may exist or develop over time in data scientists and analysts, independent of their use of interventions. 
These biases could influence their decision-making processes and perpetuate existing inequities\cite{long-term, mendelsohn2019creatures}, which may be less responsive to intervention. 
Thus recognizing the boundaries of where interventions may be effective is important for designing more ambitious interventions. 
}

\subsection{Challenge 3: Measuring Efficacy \& Boosting Adoption}
We hypothesize that a collection of complementary evaluation techniques will be needed to understand the complex interplay between system behavior and user behavior when measuring behavior change in data science.

\remove{For example, our review identifies theoretical scenarios where analysts may/may not be receptive to disruptions caused by digital interventions~\cite{dresser1985patients,gjellestad2022trust}. However once disrupted, h}
\revise{H}ow do we measure the efficacy of deployed interventions? Are the same metrics used to choose a disruptor and an intervention sufficient to understand their efficacy? It also becomes crucial to isolate whether the cause of positive behavior change is indeed the intended intervention, or attributable to some other confounding factor. \remove{Further, it is unclear how to detect lasting changes in user behavior; would repeated measures of the same heuristic over time provide sufficient information to show progress? Or do we need to measure specific long-term outcomes, such as by analyzing the comprehensiveness of project deliverables? Effective measures can help us identify when a certain intervention is no longer needed.}
\revise{Furthermore, would repeated measures of the same heuristic over time provide sufficient information to show progress? Or do we need to measure specific long-term outcomes\cite{long-term}? Incorporating these long-term fairness considerations can provide a more comprehensive view of the effectiveness of behavior change interventions. For instance, tracking the long-term outcomes of interventions on loan approval fairness can reveal whether initial improvements in fairness metrics translate into sustained equitable lending practices.}

\remove{The ability to measure intervention efficacy can also help to boost the adoption of an intervention framework, which may otherwise receive pushback as observed in other domains, \eg among pilots or medical doctors~\cite{dresser1985patients, gjellestad2022trust}. Boosting adoption also entails understanding additional factors such as possible disparity between how receptive data analysts are to positive behavior change in theory and in practice, where disruptions to their workflows might discourage them from using intervention frameworks. Future research in this space will require careful empirical validation of interventions with analysts to ensure their applicability to real-world workflows, as well as a focus on personalized interventions to account for individual differences across analysts. Complementing our call for evaluation methods, new techniques will also be needed to gauge the successful deployment and adoption of interventions in the field, which will provide helpful indicators for community-level improvements in responsible practices.}

\subsection{Challenge 4: Incentives Versus Consequences to Induce Behavior Change}

Encouraging positive behaviors or punishing negative behaviors is analogous to a carrot versus stick metaphor. 
The examples we emphasize in this paper (\eg in \autoref{sec:examples}) primarily focus on positive reinforcement (carrots). 
However, these so-called ``carrots'' are not the only way to encourage responsible data science practices. Alternatively, how to establish consequences (sticks) operationalized into interventions as a way to \textit{enforce} course correction has yet to be explored. The BCT taxonomy~\cite{michie2013behavior} identifies relevant interventions in the categories of \bct{Reward and Threat} and \bct{Scheduled Consequences} which target the \fbc{capability} of the analyst through \moa{behavioral regulation} or by changing their \moa{attitudes towards the behavior}. \remove{The negative consequences of flawed data collection, interpretation, and prediction can translate to social, psychological, financial, and even physical harm to real people.} However, it is unclear what role data science tools should play in holding analysts accountable for their contributions to irresponsible data science outcomes. For example, how do we infer the scope of an analyst's contribution to a certain outcome, positive and/or negative? Once this scope is established, how do we reason about the consequences of an analyst's contributions in relation to the final outcomes? \remove{Finally, how might data science tools be involved in this data collection and review process? Existing initiatives to promote transparency in data science practice may prove critical to addressing these problems.}


\remove{Another primary challenge of a behavior change lens on responsible data science is ensuring its ongoing relevance as our collective understanding of technical, behavioral, and psychological phenomena evolve. For example, current theories of behavior change could be subsumed by newer theories in the future. Similarly, current ethical frameworks could be replaced by new ones as we deepen our understanding of the ethical ramifications of data science work (and technology in general). Given that these theories may influence the entire intervention design process,} \remove{it may be nontrivial to replace them within implemented interventions. In the future, it would be interesting to evaluate the strengths and limitations of current theories as they pertain to data science to anticipate knowledge gaps that may be filled by new theories. With this information, we could build scaffolding into interventions for known knowledge gaps in advance, and expand on this scaffolding as new theories address them in the future. Another interesting direction is to develop modular design strategies, one example being domain-specific languages for implementing theories of behavior change, to make it easier to swap theories within the implementation of a given intervention. In this case, one could implement new theories using the language to enable automatic updates to downstream interventions. Given domain-specific languages have been developed for a wide range of scenarios in data science (\eg \cite{mcnutt2023grammar,jun2022tisane,lip62021}), we believe this to be a promising direction for future work.}

\subsection{Challenge \revise{5}: \remove{Tackling Tension between} Automated \remove{and} \revise{Versus} Behaviorally Responsible Data Science}

There exists a tension between automation and behavioral responsibility. For example, autoML techniques aim to reduce the reliance on analysts for making design decisions towards creating satisfactory models~\cite{karmaker2021automl}. While these methods reduce the analyst's time and effort in generating satisfactory models, autoML methods are poorly designed to support human oversight and agency within this process~\cite{crisan2021fits}. With a reduced ability to intervene in the model design process, the analyst's behavioral responsibilities may clash with the goals of autoML systems. \remove{For example, how can analysts ensure the model is behaving responsibly if they are minimally involved in the model development phase?} Further investigation is needed to understand how behavioral responsibility can meaningfully engage with highly automated data science tools.

\revise{\subsection{Challenge \revise{6}: Enhancing Education and Training for Data Science Practitioners} Throughout this paper, we highlight the importance of education in promoting responsible data science practices. Echoed by many prior works
in this space~\cite{attwood2019global,srikant2017introducing,brunner2016teaching}, one potential direction for the responsible data science research community is to delve deeper into developing comprehensive educational frameworks and training programs that equip the current and future data science practitioners with the necessary skills and ethical mindset to navigate complex data science environments. These programs should go beyond technical proficiency to include modules on ethical reasoning, bias detection and mitigation, and the societal impacts of data science decisions. Additionally, integrating behavior change theories into training curricula can help instill long-lasting responsible behaviors. Research should also explore innovative teaching methods, such as experiential learning\cite{sandbox}, case studies\cite{ML-tutorial}, and interactive simulations\cite{reshape}, to enhance the learning experience. 
By advancing education and training, we can prepare data scientists to not only excel technically but also to act responsibly and ethically in their professional roles.}

%% file: sections/6.conclusion.tex
\section{Conclusion}



\revise{ In this paper, we introduce the concept of behavior change interventions for data science, emphasizing that data science behaviors can be predictors of biased outcomes. Our work synthesizes a definition of responsible behaviors in data science, encompassing both human (behavioral) and system (technical) aspects. To characterize interventions within data science contexts, we illustrate how existing psychological models can inform the design of behavior change interventions. We operationalize these theories through a four-step framework to help design effective behavior change interventions, which includes (1) identifying problematic and target behaviors, (2) identifying factors affecting problematic behaviors, (3) understanding and employing appropriate Mechanisms of Action, and (4) envisioning potential interventions using Behavior Change Techniques. To inspire the design of practical interventions, we present concrete examples that encourage socially responsible behaviors within the data science context. We conclude by describing the open challenges uncovered by this vision paper and call on our community to explore this emerging research area of behavior change interventions for responsible data science, thereby promoting ethical and socially responsible practices in the field.}